\documentclass[11pt,a4paper]{article}
\usepackage{jcappub}

\usepackage{bm}
\usepackage{epsfig}
\usepackage{graphicx}
\usepackage{amsmath}
\usepackage{amssymb}
\usepackage{amsbsy}
\usepackage{color}
\usepackage{subfigure}
\usepackage{slashed}
\usepackage{afterpage}
\usepackage{psfrag}
\usepackage{axodraw}

\newcommand{\be}{\begin{equation}}
\newcommand{\ee}{\end{equation}}
\newcommand{\bea}{\begin{eqnarray}}
\newcommand{\eea}{\end{eqnarray}}

\definecolor{gre}{rgb}{0,0.4,0.3}

\begin{document}
\subheader{\hfill MPP-2012-129}

\title{CAST constraints on the axion-electron coupling}

\author[a]{K.~Barth,}
\author[b]{A.~Belov,}
\author[c,2]{B.~Beltran,}
\author[d]{H.~Br\"auninger,}
\author[c]{J.~M.~Carmona,}
\author[e]{J.~I.~Collar,}
\author[c,j,f]{T.~Dafni,}
\author[a]{M.~Davenport,}
\author[a,3]{L.~Di~Lella,}
\author[g]{C.~Eleftheriadis,}
\author[d]{J.~Englhauser,}
\author[h]{G.~Fanourakis,}
\author[j]{E.~Ferrer-Ribas,}
\author[k]{H.~Fischer,}
\author[k]{J.~Franz,}
\author[d]{P. Friedrich,}
\author[c,j]{J.~Gal\'an,}
\author[c]{J.~A.~Garc\'ia,}
\author[h]{T.~Geralis,}
\author[j]{I.~Giomataris,}
\author[b]{S.~Gninenko,}
\author[c,4]{H.~G\'omez,}
\author[l]{M.~D.~Hasinoff,}
\author[k,5]{F.~H.~Heinsius,}
\author[f]{D.~H.~H.~Hoffmann,}
\author[c,a,j]{I.~G.~Irastorza,}
\author[m]{J.~Jacoby,}
\author[n]{K.~Jakov\v{c}i\'{c},}
\author[k,6]{D.~Kang,}
\author[k]{K.~K\"onigsmann,}
\author[o]{R.~Kotthaus,}
\author[h,7]{K.~Kousouris,}
\author[n]{M.~Kr\v{c}mar,}
\author[d,f,8]{M.~Kuster,}
\author[n]{B.~Laki\'{c},}
\author[g]{A.~Liolios,}
\author[n]{A.~Ljubi\v{c}i\'{c},}
\author[p,9]{G.~Lutz,}
\author[c]{G.~Luz\'on,}
\author[e]{D.~W.~Miller,}
\author[a,j]{T.~Papaevangelou,}
\author[q]{M.~J.~Pivovaroff,}
\author[o]{G.~Raffelt,}
\author[o,r]{J.~Redondo,}
\author[f]{H.~Riege,}
\author[c]{A.~Rodr\'iguez,}
\author[q,a,c]{J.~Ruz,}
\author[g]{I.~Savvidis,}
\author[s]{Y.~Semertzidis,}
\author[a]{L.~Stewart,}
\author[q,10]{K.~Van~Bibber,}
\author[e,11]{J.~D.~Vieira,}
\author[c]{J.~A.~Villar,}
\author[q,k]{J.~K.~Vogel,}
\author[a]{L.~Walckiers,}
\author[f,a,t]{K.~Zioutas}

\affiliation[a]{\small European Organization for Nuclear Research (CERN), Gen\`eve, Switzerland}
\affiliation[b]{\small Institute for Nuclear Research (INR), Russian Academy of Sciences, Moscow, Russia}
\affiliation[c]{\small Laboratorio de F\'{\i}sica Nuclear y Altas Energ\'{\i}as, Universidad de Zaragoza, Zaragoza, Spain}
\affiliation[d]{\small Max-Planck-Institut f\"{u}r extraterrestrische Physik, Garching, Germany}
\affiliation[e]{\small Enrico Fermi Institute and KICP, University of Chicago, Chicago, IL, USA}
\affiliation[f]{\small Technische Universit\"{a}t Darmstadt, IKP, Darmstadt, Germany}
\affiliation[g]{\small Aristoteles University of Thessaloniki, Thessaloniki, Greece}
\affiliation[h]{\small National Center for Scientific Research ``Demokritos'', Athens, Greece}
\affiliation[j]{\small IRFU, Centre d'\'Etudes Nucl\'eaires de Saclay (CEA-Saclay), Gif-sur-Yvette, France}
\affiliation[k]{\small Albert-Ludwigs-Universit\"{a}t Freiburg, Freiburg, Germany}
\affiliation[l]{\small Department of Physics and Astronomy, University of British Columbia, Vancouver, Canada}
\affiliation[m]{\small J.~W.~Goethe-Universit\"at, Institut f\"ur Angewandte Physik, Frankfurt am Main, Germany}
\affiliation[n]{\small Rudjer Bo\v{s}kovi\'{c} Institute, Zagreb, Croatia}
\affiliation[o]{\small Max-Planck-Institut f\"{u}r Physik, Munich, Germany}
\affiliation[p]{\small MPI Halbleiterlabor, M\"unchen, Germany}
\affiliation[q]{\small Lawrence Livermore National Laboratory, Livermore, CA, USA}
\affiliation[r]{\small Arnold Sommerfeld Center, Ludwig-Maximilians-Universit\"at, Munich, Germany}
\affiliation[s]{\small Brookhaven National Laboratory, Brookhaven, USA}
\affiliation[t]{\small Physics Department, University of Patras, Patras, Greece}

\footnotetext[1]{Present address: Institute de Physique Nucl\'eaire, Lyon, France}
\footnotetext[2]{Present address: Department of Physics, Queens University, Kingston, Ontario}
\footnotetext[3]{Present address: Sezione di Pisa, IT}
\footnotetext[4]{Present address: Laboratoire de l'Acc\`{e}l\`{e}rateur Lin\`{e}aire, Centre Scientifique d'Orsay, 91898 Orsay, France}
\footnotetext[5]{Present address: Institut f\"{u}r Experimentalphysik, Ruhr-Universit\"{a}t Bochum, Bochum, Germany}
\footnotetext[6]{Present address: Karlsruhe Insitute of Technology, Germany}
\footnotetext[7]{Present addtess: European Organization for Nuclear Research (CERN), Gen\`eve, Switzerland}
\footnotetext[8]{Present address: European XFEL GmbH, Notkestrasse 85, 22607 Hamburg, Germany}
\footnotetext[9]{Present address: PNSensor GmbH, M\"unchen, Germany}
\footnotetext[10]{Present address: Department of Nuclear Engineering. University of California Berkeley, USA}
\footnotetext[11]{Present address: California Institute of Technology, USA}
\emailAdd{Jaime.Ruz@cern.ch}
\emailAdd{Julia.Vogel@cern.ch}
\emailAdd{redondo@mpp.mpg.de}

\abstract{In non-hadronic axion models, which have a tree-level
axion-electron interaction, the Sun produces a strong axion flux by
bremsstrahlung, Compton scattering, and axio-recombination, the
``BCA processes.'' Based on a new calculation of this flux,
including for the first time axio-recombination, we derive limits on
the axion-electron Yukawa coupling $g_{ae}$ and axion-photon
interaction strength $g_{a\gamma}$ using the CAST phase-I data
(vacuum phase). For $m_a\lesssim 10~\rm{meV/c^{2}}$ we find
$g_{a\gamma}\,g_{ae}< 8.1\times 10^{-23}\, {\rm GeV}^{-1}$ at
95\%~CL. We stress that a next-generation axion helioscope such as
the proposed IAXO could push this sensitivity into a range beyond
stellar energy-loss limits and test the hypothesis that
white-dwarf cooling is dominated by axion emission.}

\keywords{axions, helioscopes, sun, magnetic fields, white dwarfs}
\arxivnumber{1302.6283}

\maketitle

\section{\label{intro}Introduction and main results}

The CERN Axion Solar Telescope (CAST)~\cite{Zioutas:1998cc} is a
helioscope experiment aiming at the detection of axions and
axion-like particles (ALPs) emitted from the Sun. The detection
principle is based on the axion\footnote{Unless otherwise noted, the
term ``axion'' henceforth includes both QCD axions and more general
ALPs.} coupling to two photons, which triggers their conversion into
photons of the same energy as they propagate through a transverse
magnetic field~\cite{Sikivie:1983ip,Raffelt:1987im}. CAST has tracked
the Sun in three different campaigns
(2003--04~\cite{Zioutas:2004hi,Andriamonje:2007ew},
2005--2006~\cite{Arik:2008mq} and \hbox{2008 \cite{Arik:2011rx}}) with
a 9.26~m long, 9~Tesla strong, decommissioned LHC dipole test magnet
while measuring the flux of X-rays at the exits of both bores with
four different low-background detectors.  No excess counts were
observed over the expected backgrounds, thus constraining the axion
parameters, notably mass and couplings to
photons~\cite{Zioutas:2004hi,Andriamonje:2007ew,Arik:2008mq,Arik:2011rx}
and nucleons~\cite{Andriamonje:2009ar,Andriamonje:2009dx}. Such
constraints require knowledge of the flux of axions emitted from the Sun
which can be computed very precisely  because the solar interior is a
tractable weakly-coupled plasma.

The physics case of CAST was mainly focused on hadronic
axions~\cite{Kim:1979if,Shifman:1979if} which were appealing as hot
dark matter candidates~\cite{Moroi:1998qs}. Hadronic axion models are
minimal in that the generic axion interactions with hadrons and
photons derive from mixing with the pseudoscalar mesons $\pi^0$,
$\eta$ and $\eta'$. The interactions with leptons arise at loop
level~\cite{Srednicki:1985xd} and are usually irrelevant. In hadronic
models, the bulk of the solar axion flux comes from Primakoff
production $\gamma+Q\to a+Q$~\cite{Dicus:1978fp,Raffelt:1985nk,Raffelt:1987np}, where $Q$ is
any charged particle.

Recently, non-minimal axion models are receiving increasing
attention~\cite{Svrcek:2006yi,Arvanitaki:2009fg,Acharya:2010zx,Giudice:2012zp,Cicoli:2012sz,Redi:2012ad,Ringwald:2012hr,Jaeckel:2012yz,Hertzberg:2012zc,Chatzistavrakidis:2012bb}.
From the theoretical point of view, axions are nowadays known to
arise naturally in many extensions of the standard model that pursue
some sort of unification, such as Grand Unified Theories or string
theory. Indeed, the original hadronic KSVZ
axion~\cite{Kim:1979if,Shifman:1979if} can be regarded as an
exemplary toy model that contains only the essential ingredients to
solve the strong CP problem~\cite{Peccei:2006as}, but often axions
arising in completions of the standard model are not minimal in this
sense.

Non-hadronic axion models such as that of
DFSZ~\cite{Zhitnitsky:1980tq,Dine:1981rt} have different and very
interesting phenomenological consequences. Notably, they couple to
electrons at tree level, and this opens axion-production channels in
stars which are much more effective than the Primakoff process:
electron-ion bremsstrahlung ($e+I\to
e+I+a$)~\cite{Raffelt:1985nk,Krauss:1984gm}, electron-electron
bremsstrahlung ($e+e\to e+e+a$)~\cite{Raffelt:1985nk}, Compton
($\gamma+e\to e+a$)~\cite{Fukugita:1982ep,Fukugita:1982gn},
axio-recombination ($e+I\to
I^-+a$)~\cite{Dimopoulos:1986kc,Dimopoulos:1986mi,Derevianko:2010kz,Redonbination}
and, to a lesser extent, axio-deexcitation of ions ($I^*\to I+a$).
Henceforth we shall refer to this set of reactions as BCA for its
most relevant contributions from bremsstrahlung, Compton, and
axio-recombination. Indeed, axions with $g_{ae}\sim{}
10^{-13}$ might explain the longstanding anomaly in the cooling of
white dwarfs (WD)~\cite{Isern:1992}, recently reinforced by updated
studies of the period decrease of the pulsating white dwarfs
G117-B15A~\cite{Isern:2010wz,Corsico:2012ki} and
R548~\cite{Corsico:2012sh} and the WD luminosity
function~\cite{Isern:2008nt,Isern:2008fs,Isern:2012ef,Melendez:2012iq}.
One should note that this value is somewhat challenged by the
constraint $g_{\rm ae}< 2.5\times 10^{-13}$ imposed by the evolution
of red giant stars in globular clusters
\cite{Raffelt:1989xu,Raffelt:1994ry}. These values of $g_{ae}$ imply
a DFSZ-axion decay constant $f_{a}\sim10^9$ GeV (corresponding to an
axion mass $m_{a} \sim$ meV). Such meV-mass axions have a wealth of
other interesting phenomenological implications in the context of
astrophysics, like the formation of a cosmic diffuse background of
axions from core collapse supernova explosions~\cite{Raffelt:2011ft}
or neutron star cooling~\cite{Umeda:1997da,Keller:2012yr}. In
cosmology, the decay of relic axionic strings and domain walls
produces a relevant cold dark matter
population~\cite{Hiramatsu:2012gg}.

\begin{figure}[tbp]
   \centering
   \includegraphics[width=8cm]{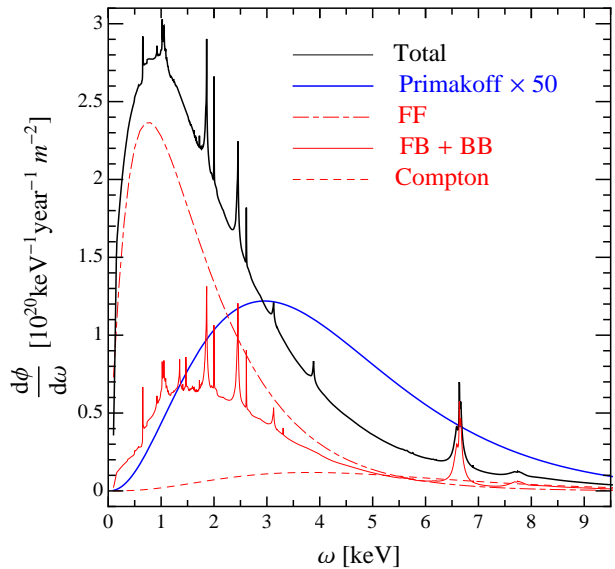}
   \vspace{-0.2cm}
   \caption{Solar axion flux on Earth for a typical DFSZ model
   with interaction strength to photons $g_{a\gamma}= 10^{-12}$~GeV$^{-1}$
   and electrons $g_{ae}= 10^{-13}$, corresponding to $f_a=0.85\times 10^{9}$~GeV
   \cite{Redonbination}.
   The blue line corresponds to the Primakoff flux and the red lines show the 
   different components of the BCA flux: FF = free-free (bremsstrahlung), FB = free-bound (axio-recombination), and
   BB = bound-bound (axio-deexcitation). The black line is the total flux. }
   \vspace{-0.2cm}
   \label{fig:flux}
\end{figure}

Still, besides new studies of red giant evolution --- currently
underway~\cite{RafRedViauxinprep} --- or features in massive star
evolution such as~\cite{Friedland:2012hj} we know of no other way to
test the WD cooling hypothesis that could rely \emph{solely} on the
axion-electron coupling. It appears thus that to assess the WD
cooling hypothesis we must study processes involving other axion
couplings as well. 
In this paper we will make use of the coupling to two photons.
As already mentioned, the solar BCA flux of non-hadronic axions is
generically much larger than that of hadronic models for the same
value of $f_a$ and moreover, it has a different spectrum
(see~figure~\ref{fig:flux}). Therefore, helioscopes are appealing to
search for non-hadronic axions. Indeed, some of us have recently
shown that a next generation helioscope~\cite{Irastorza:2011gs}, such
as the proposed International AXion Observatory
(IAXO)~\cite{Irastorza:2012qf} can test the WD cooling hypothesis
down to very small couplings $g_{ae}\sim 10^{-13}$.

In the present paper we analyze the CAST data in search of
non-hadronic axions and set new upper bounds on $g_{ae}\times
g_{a\gamma}$, the product of the electron coupling (responsible for
the production in the Sun) and the two-photon coupling (responsible
for the detection in CAST). 

\begin{figure}[tbp]
   \centering
   \includegraphics[width=8cm]{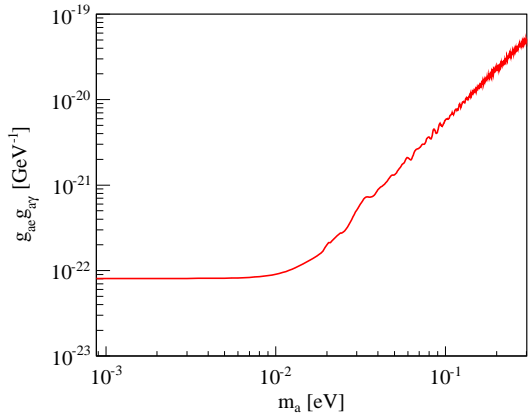}
   \vspace{-0.2cm}
\caption{CAST constraints on $g_{ae}\times g_{a\gamma}$ as a function of $m_a$,
   assuming the solar
   emission is dominated by the BCA reactions which involve only the electron coupling $g_{ae}$.}
   \vspace{-0.2cm}
   \label{fig:gaegag}
\end{figure}

Figure~\ref{fig:gaegag} shows our results when we assume that the
Primakoff emission from the Sun is subdominant and therefore the
solar flux is caused by the BCA reactions alone. Our analysis of
CAST data then constrains
\begin{equation}
\label{eq:gaegag}
g_{a\gamma}\times g_{ae}< 8.1\times 10^{-23}\, {\rm GeV}^{-1} \quad (95\%~\rm CL)
\end{equation}
at low masses $m_a\lesssim 10$ meV --- where the probability of
axion-photon conversion in CAST becomes independent of the mass
--- and worsens as $1/m_a^2$ for higher masses.

\begin{figure}[tbp]
   \centering
   \includegraphics[width=7.5cm]{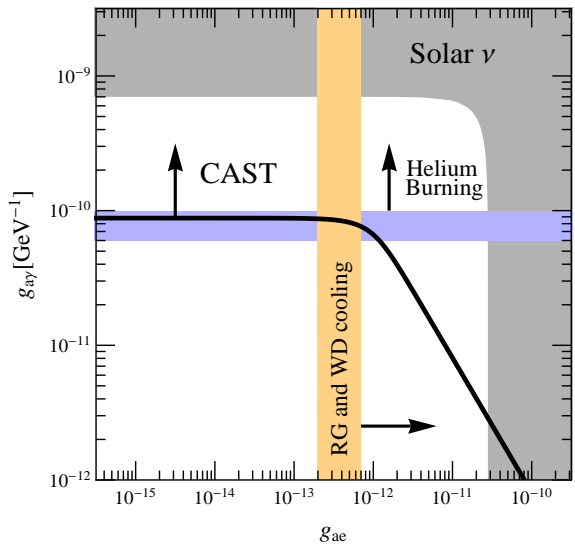}  
   \caption{Constraints on $g_{ae}$ and $g_{a\gamma}$ for $m_a\lesssim10$~meV.
   The region above the thick black line is excluded by CAST.
   The gray region is excluded by solar neutrino measurements.
   In the vertical orange band, axion emission strongly
   affects white dwarf cooling and the evolution of low-mass
   red giants; parameters to the right of this band are excluded.
    Likewise, helium-burning stars would be perceptibly affected in
    the horizontal blue band; parameters above it are excluded.}
   \label{fig:gaegag2}
\end{figure}

If we also include the Primakoff flux (which is unavoidable because
it is produced by the same coupling $g_{a\gamma}$ involved in the
CAST detection), the signal at CAST depends independently on three
parameters: $g_{ae}$, $g_{a\gamma}$ and $m_a$. However, for
$m_a\lesssim 10$~meV the detection is independent of mass and we can
plot our results in the $g_{ae}$--$g_{a\gamma}$ parameter space. In
this low-mass range, phase-I of CAST gives the strongest constraints
and thus we have focused only on this data set. Our analysis, based 
on a two-free parameter likelihood method is able to exclude
the region above the thick black line in figure~\ref{fig:gaegag2}.
For very small values of $g_{ae}\lesssim 10^{-12}$, the BCA flux is
negligible and the CAST bound smoothly becomes $g_{a\gamma}<0.88
\times 10^{-10}$ GeV$^{-1}$ as found in our previous
study~\cite{Andriamonje:2007ew} where only Primakoff emission was
assumed. However, for larger values of $g_{ae}$ the BCA flux becomes
dominant and we recover equation~\ref{eq:gaegag}.

Note that our bound relies on a simple calculation of the solar axion
flux, for which we have taken a solar model unperturbed by axion
emission. If $g_{ae}$ or $g_{a\gamma}$ are very large, the large
axion flux requires a modified internal structure of the Sun with
larger nuclear reaction rates and higher temperature of the core. The
most stringent constraint derives from the agreement between the
predicted and observed solar boron neutrino
flux~\cite{Gondolo:2008dd}, excluding the gray region depicted in
figure~\ref{fig:gaegag2} (labeled Solar $\nu$). Thus our bound is
completely self-consistent up to $g_{ae}=3\times 10^{-11}$, in
contrast to those solar axion searches utilising Bragg
scattering~\cite{Avignone:1997th,Morales:2001we,Bernabei:2001ny,Ahmed:2009ht}
which have more limited sensitivity~\cite{Cebrian:1998mu}, as well as to some 
other searches relying solely on $g_{ae}$ coupling~\cite{Kekez, Derbin, Abe}.

In order to put our results into context we also show in
figure~\ref{fig:gaegag2} two color bands representing parameters
where axion emission would have a strong impact on stellar evolution.
In the vertical orange band of $g_{ae}$ values, axion emission would
strongly affect WD
cooling~\cite{Isern:1992,Isern:2010wz,Corsico:2012ki,Corsico:2012sh,Isern:2008nt,Isern:2008fs,Isern:2012ef,Melendez:2012iq}
and delay helium ignition in low-mass red
giants~\cite{Raffelt:1989xu,Raffelt:1994ry}. The exact range of
$g_{ae}$ values that is consistently ruled in or ruled out by these
arguments remains to be studied in detail, but for sure parameters to
the right of this band are excluded. Within the horizontal blue band,
axion Primakoff emission would strongly affect stars in the
helium-burning phase. The upper edge of this band corresponds to the
traditional horizontal-branch star limit, the remaining range
represents a new argument concerning the blue-loop suppression during
the helium-burning phase of massive stars~\cite{Friedland:2012hj}.

The orange band cuts the CAST constraint in its horizontal part which
corresponds to the Primakoff flux dominating the solar flux, but very
close to the values $g_{ae}\sim 10^{-12}$ where the BCA flux starts
to dominate. Therefore, CAST cannot shed any further light on the WD
cooling hypothesis. However, a next-generation helioscope such as
IAXO with its improved sensitivity to $g_{a\gamma}$ will also benefit
from the large BCA-emitted flux and will improve over the RG bound in
part of the parameter space. In principle, the WD cooling
hypothesis is then testable in a laboratory experiment.

After having presented our results and main messages, the rest of the
paper is devoted to elaborate on our definitions, assumptions, and
analysis method. In section~\ref{sec:theory} we give a brief account
of axion theory, we further examine the implications of our findings,
and finally describe the solar axion flux, and in
section~\ref{sec:experiment} we present our new analysis after a
summary of the experimental setup of CAST phase-I and its results.

\section{Properties of axions and axion-like particles}\label{sec:theory}

For the purpose of the present paper we can parametrize an axion
model with the lagrangian density
\begin{equation}
{\cal L} = \frac{1}{2}(\partial_\mu a) (\partial^\mu a) - \frac{1}{2}m_a^2 a^2
-\frac{g_{a\gamma}}{4}F_{\mu\nu}\tilde{F}^{\mu\nu}a
-g_{a e}\frac{\partial_{\mu}a}{2m_e}\overline{\psi}_{e}\gamma_{5}\gamma^{\mu}\psi_{e}\,,
\end{equation}
where $a$ is the axion field, $m_a$ its mass, $F_{\mu\nu}$ and
$\tilde{F}^{\mu\nu}$ are the electromagnetic field tensor and its
dual, $m_e$ the electron mass, and $\psi_e$ the electron field. The
coupling constant $g_{a\gamma}$ has units of energy$^{-1}$ while
$g_{a e}$ is a dimensionless Yukawa coupling. 
Particles featuring
this type of lagrangians are often called \emph{axion-like particles} (ALPs) and
they appear as pseudo-Nambu-Goldstone bosons
(pNGB), associated with a global shift symmetry
$a\to a+{\rm const.}$ which is spontaneously broken at some
high energy scale $f_a$ (sometimes called ALP decay constant) 
or stringy axions where $f_a$ corresponds to 
the string scale $M_s$. The shift symmetry is explicitly broken by
some perturbing dynamics responsible for the mass term.

In 1977 Peccei and Quinn proposed one such symmetry to solve the
strong CP problem \cite{Peccei:1977hh,Peccei:1977ur} with the
additional condition that it should be color anomalous. The
resulting pNGB was called the axion~\cite{Weinberg:1977ma,Wilczek:1977pj}. 
The axion receives its mass from
chiral symmetry breaking after mixing with the pseudoscalar mesons
through the color anomaly, its magnitude being
\begin{equation}
\label{eq:amass}
m_{a}= \frac{\sqrt{z}}{1+z}\frac{m_{\pi} f_{\pi}}{f_{a}}\simeq 6\,{\rm meV}\frac{10^{9}{\rm GeV}}{f_a},
\end{equation}
where $m_{\pi}$ is the neutral pion mass and $f_\pi$ is the pion
decay constant. For the ratio $z=m_{u}/m_{d}$ of up to down quark
masses we use the canonical value $z\sim 0.56$ although the allowed
range is $z=0.35$--0.60~\cite{PDG2010}, however this only leads to
minor uncertainties in our context.

The axion has a model-independent contribution to its two-photon
coupling coming from the above-mentioned mixing with mesons and can
also have a model-dependent part if the PQ symmetry has the
electromagnetic anomaly. The two contributions sum to
\begin{equation}
g_{a\gamma} = \frac{\alpha}{2\pi f_a}\left(\frac{E}{N}-\frac{2}{3}\frac{4+z}{1+z}\right)
\simeq
\frac{\alpha}{2\pi f_a}\left(\frac{E}{N}-1.92\right)\,,
\end{equation}
where $\alpha$ is the fine-structure constant and $E/N$ the ratio of
the electromagnetic and color anomalies of the PQ symmetry.

The coupling to electrons has a model-dependent contribution
proportional to an $O(1)$ coefficient $X_e$ arising only in
non-hadronic axion models and a very small model-independent one
induced at one-loop via the photon coupling, 
\begin{equation}
g_{ae} = X_e \frac{m_e}{f_a}+\frac{3\alpha^2}{4\pi}\frac{m_e}{f_a}\left(\frac{E}{N}\log\frac{f_a}{m_e}-1.92\, \log\frac{\Lambda}{m_e}\right)\, ,
\end{equation}
where $\Lambda$ is an energy scale close to the QCD confinement
scale.

For a generic ALP, $\phi$, we expect similar equations for $g_{\phi e}$ and
$g_{\phi \gamma}$ as for the axion, of 
course after removing the terms coming from axion-meson mixing,
(the terms involving $z$) and changing $f_a$ for the ALP decay constant, $f_\phi$. 
However, the ALP mass is then completely unrelated to the couplings.

\section{Expected counting rate}\label{sec:events}

\subsection{Solar flux}

\begin{figure}[!hqtbp] 
    \centering
    \includegraphics[width=11cm]{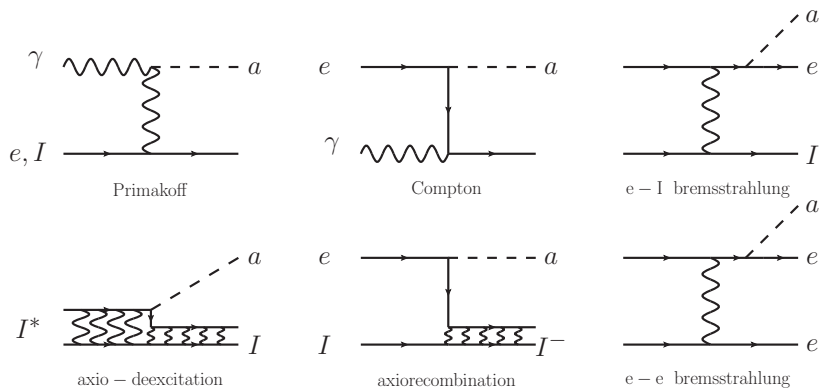}
    \caption{Some Feynman diagrams for the most relevant solar axion emission reactions included in this work.}
    \label{fig:diagrams}
 \end{figure}

Based on the different axion interactions, different processes
contribute to the production of the solar flux of these particles
(see figure \ref{fig:diagrams}). The most important processes are:
\begin{itemize}
\item Primakoff effect: $\gamma + Q \rightarrow Q + a$
\item Compton scattering (photo production): $\gamma + e
    \rightarrow  e + a$
\item Electron-Ion bremsstrahlung (free-free transition): $e+I\rightarrow  e + I+ 
    a$
\item Electron-electron bremsstrahlung: $e+e\rightarrow
    e+e+a$
\item Axio-recombination (free-bound transition):
    $e+I\rightarrow I^-+a$
\item Axio-deexcitation (bound-bound transition): $I^*\rightarrow
    I+a$
\end{itemize}
where $Q$ stands for any charged particle in the solar plasma, $e$ for electrons,
$I$ for ions and $I^*$ for their excited states. The Primakoff process depends on 
the two-photon coupling and dominates when the coupling to electrons is absent at 
tree-level. When this is not the case the BCA processes dominate: bremsstrahlung
on hydrogen and helium nuclei dominates the emission of low-energy
axions, axio-recombination of metals (mostly O, Ne, Si, S and Fe)
contributes sizably at intermediate energies and Compton takes over
at higher energies. The contribution of axio-deexcitation is
dominated by Lyman transitions (mostly Ly$-\alpha$) and is
significant only in the case of iron which dominates the axion flux
around $\sim 6.5$~keV.
Following Ref.~\cite{Raffelt:1985nk} and integrating the emission
rates over a solar model~\cite{TurckChieze:2001ye}, the following
fits for the axion fluxes at Earth can be obtained~\cite{Irastorza:2011gs} 
(in units of  $\rm{m^{-2}~{\rm year}^{-1}~keV^{-1}}$)
\begin{eqnarray}
\frac{{\rm d}\Phi_a}{{\rm d}\omega}\Big \vert_{P} &=& 2.0\times 10^{18}\,\Big( \frac{g_{a\gamma}}{10^{-12}{\rm GeV}^{-1}} \Big)^{2} \, \omega^{2.450}\,e^{-0.829\,\omega} \\
\frac{{\rm d}\Phi_a}{{\rm d}\omega}\Big \vert_{C} &=&
4.2\times 10^{18}\,\Big( \frac{g_{a e}}{10^{-13}} \Big)^{2} \,\omega^{2.987}\,e^{-0.776\,\omega} \\
\frac{{\rm d}\Phi_a}{{\rm d}\omega}\Big \vert_{B} &=& 8.3\times 10^{20}\,\Big( \frac{g_{a e}}{10^{-13}} \Big)^{2} \,\frac{\omega}{1+0.667\,\omega^{1.278}}\,e^{-0.77\,\omega}
\end{eqnarray}
where $\omega$ is the axion energy in keV, while P, C and B stand for
Primakoff, Compton, and bremsstrahlung, respectively. In particular,
the bremsstrahlung flux includes both electron-nucleus (only H and
He) and electron-electron contributions. As a novelty, we include the
emission of axions in the electron capture by an ion, dubbed
``axio-recombination'' (free-bound transition) and atomic
``axio-deexcitation" (bound-bound transition). Originally, the former
process was estimated to be subdominant in the
Sun~\cite{Dimopoulos:1986kc,Dimopoulos:1986mi}. However, a new
calculation to be presented elsewhere~\cite{Redonbination}, including
the missing factor of 2 in the cross-section pointed out
in~\cite{Derevianko:2010kz} and captures in higher shells than the
K-shell, shows that these processes are significant and increase the
total flux by a factor of order~1 (figure~\ref{fig:flux}).
Unfortunately, the kinematic edges and the narrow lines from
bound-bound transitions prevent us from providing a simple fitting
formula.

\subsection{Helioscope event number}

The expected number of photons $\mathcal{N}_{\gamma}$ from axion
conversion in a given detector is obtained by integrating the product
of the differential axion flux, conversion probability and detection efficiency over the total range of energies
\begin{equation}\label{ngamma}
\mathcal{N}_{\gamma} =  \int_{\omega_{0}} ^{\omega_{f}}\mathrm{d}\omega\, \Big( \frac{\mathrm{d}\Phi_a}{\mathrm{d}\omega}\Big)_{\rm total}\, \mathcal{P}_{a\rightarrow \gamma}\, \epsilon \, S\, t
\end{equation}
where $S$ is the detection area perpendicular to the flux of axions,
$t$ is the exposure time, and $\epsilon$ the detection efficiency.
The axion-photon conversion probability in a transverse homogeneous
magnetic field $B$ over distance $L$ is
\begin{equation}\label{prob}
P_{a \rightarrow \gamma} = \left(\frac{g_{a\gamma}B L}{2}\right)^2{\rm sinc}^2\left(\frac{q L}{2}\right)\,,
\end{equation}
where sinc$\,x=(\sin x)/x$ and the momentum transfer provided by the
magnetic field is $q=m_a^2/2\omega$. Coherent $a$-$\gamma$
conversion along the full magnetic length happens when
$m_a^2<4\omega/L$, i.e.\ when the momentum transfer is smaller
than about $1/L$, where sinc$(qL/2)\to 1$. For larger masses, the
conversion is not coherent\footnote{We only use data from the
CAST vacuum phase where photon refraction by the residual gas can be
neglected due to the high vacuum
conditions~\cite{Andriamonje:2007ew}.} and the probability gets suppressed by a
factor $\sim(4\omega/m_a^2 L)^2$. This factor is responsible for
the degradation of our bound for $m_a\gtrsim10$~meV seen in
figure~\ref{fig:gaegag}.

\section{CAST experiment and analysis}\label{sec:experiment}

The most sensitive helioscope to date is the CERN Axion Solar
Telescope (CAST), which makes use of a prototype superconducting LHC
dipole magnet providing a magnetic field of up to $\rm{9\,T}$. CAST
is able to follow the Sun twice a day during sunrise and sunset
for a total of $\rm{3\,h}$ per day. At both ends of the $\rm{9.26\,m}$
long magnet, X-ray detectors~\cite{Kuster:2007ue,Abbon:2007ug,Autiero:2007uf} have been
mounted to search for photons from Primakoff conversion.

CAST began its operation in 2003 and, after two years of data
taking, determined an upper limit on $g_{a \gamma} \lesssim 0.88
\times 10^{-10}\,\rm{GeV^{-1}}$ at $95\,\rm{\%}$ CL for $m_{a}\leq
0.02\,\rm{eV}$~\cite{Zioutas:2004hi,Andriamonje:2007ew} in the
hadronic axion scenario. To extend the experimental sensitivity to
larger axion masses, the conversion region of CAST was filled with a
suitable buffer gas~\cite{vanBibber:1988ge} providing the photons
with an effective mass, yielding upper limits on
$g_{a\gamma}\lesssim 2.3\times 10^{-10}\rm{GeV^{-1}}$ for
$0.02\leq m_{a}\leq 0.65\,\rm{eV}$~\cite{Arik:2008mq,Arik:2011rx}.
Higher axion masses have also been studied and future publications
will report on these masses which cover a range up to
$1.18\,\rm{eV}$.

\subsection{The X-ray telescope of CAST}
The most sensitive detector system operative at CAST phase-I was 
the X-ray telescope \cite{Autiero:2007uf}, a combination of X-ray mirror
optics~\cite{Friedrich} and a Charge-Coupled Device
(CCD)~\cite{Strder} located in the focal plane of the mirror optics.
Both instruments were originally built for satellite space missions,
and together they increase the axion discovery potential
significantly along with providing excellent imaging capability. The
implementation of the X-ray mirror optics suppresses background by a
factor of $\sim 155$, since photons are focused from the magnet
aperture area of about $14.5\,\rm{cm^{2}}$ to a spot of roughly
$9.3\,\rm{mm^{2}}$ on the CCD chip. One of the resulting advantages
of focusing optics is the possibility to measure background and
signal simultaneously.

\subsection{Data taking}
The CCD detector showed a stable performance over the entire 2004
running period. For our analysis, we have used a total of
$197\,\rm{h}$ of tracking data (i.e.\  magnet pointing to the Sun)
and $1890\,\rm{h}$ of background data taken from the same area
during non-tracking periods. For a detailed description of the X-ray
telescope design, its performance, and background systematics during
the 2004 data taking period we refer to Ref.~\cite{Kuster:2007ue}.
Since no significant  signal over background was detected with
the X-ray telescope during the data taking period of 2004, upper
limits for the non-hadronic axions were derived for this CAST
detector system.

\begin{figure} [!hqt]
\begin{center}
\includegraphics[width = 0.6\textwidth]{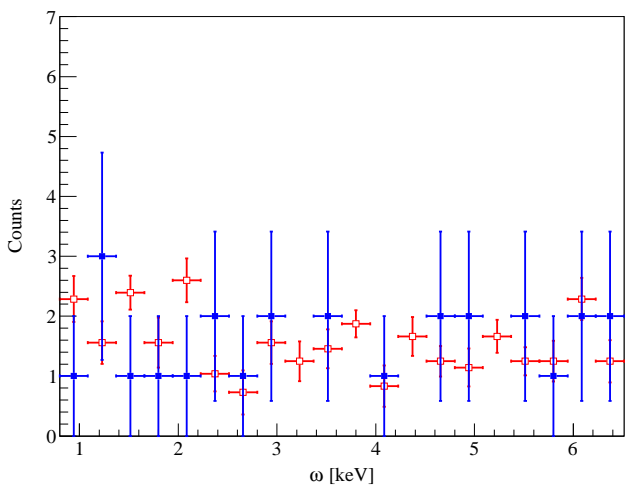}
\tiny\caption{CCD spectra for tracking (blue) and background (red) runs during 2004. In both cases, the error bars 
represent the statistical uncertainty of the measurement. Please notice that the background spectrum has been renormalized 
to the tracking time of $197~\rm{hours}$.\label{figure_spc}}
\end{center}
\end{figure}

In order to minimize the influence of the Cu-K$_{\alpha}$
fluorescence line (at $8\,\rm{keV}$ originating from the cooling
mask of the detector \cite{Kuster:2007ue}), we restricted our
analysis to the energy range between $0.8$ and $6.8\,\rm{keV}$. In
total, we observe 26 counts in this energy range inside the
signal-spot area during axion sensitive conditions. The background,
defined by the data taken from the same CCD spot region
during non-tracking periods, has been acquired under the same operating
conditions. The spectral distribution of the observed events during
tracking and non-tracking times with the CCD detector is shown in
figure~\ref{figure_spc}.

\subsection{Spectral fitting results}
The resulting low counting statistics required the use of a
maximum-likelihood method to determine an upper limit on
$g_{ae}^{2}g_{a\gamma}^{2}$. The likelihood function used is based
on a Poissonian p.d.f., the binned likelihood
\begin{equation}\label{likelihood1}
\mathcal{L} = \prod_{j}^{n}\frac{e^{-\lambda_{j}}\lambda_{j}^{t_{j}}}{t_{j}!},
\end{equation}
where $n=20$ is the number of spectral bins, $t_{j}$ the number of
observed counts in tracking, and $\lambda_{j}$ the value of the mean
in the $j$-th bin, respectively. The fit function,
$\lambda_{j}=\sigma_{j}+b_{j}$ is used, where $b_{j}$ is the
measured background and $\sigma_{j}\propto
g_{ae}^{2}g_{a\gamma}^{2}$ is the expected number of counts in the $j^{th}$ energy bin from 
axion-to-photon conversion. The best estimate for $g_{ae}^{2}g_{a\gamma}^{2}$ is obtained by minimizing
 $\chi^{2}=-2\ln{\mathcal{L}}$. The validity of the $\chi^{2}$-interpretation in our case, as well as 
the negligible influence of the statistical uncertainty of the background on the final result have been 
verified with a Monte Carlo model by means of the generation of pseudo-data sets.

\begin{figure} [!hqt]
\begin{center}
\includegraphics[width = 0.6\textwidth]{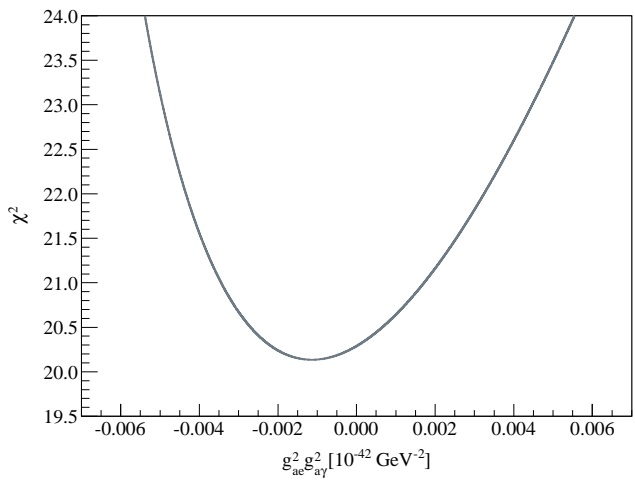} 
\tiny\caption{$\chi^{2}$ as function of $g_{ae}^{2}g_{a\gamma}^{2}$ for an axion mass of $1\,\rm{meV}$. The mimimum of the $\chi^{2}$
distribution, $-1.136\times 10^{-45}\,\rm{GeV^{-2}}$, is the most probable value of $g_{ae}^{2}g_{a\gamma}^{2}$. \label{figure_chi}}
\end{center}
\end{figure}

We compared the result derived with the maximum-likelihood defined
in equation \ref{likelihood1} with a maximum-likelihood technique
based on an unbinned maximum-likelihood estimator that divides the
exposure time in sufficiently small time fragments so that either
one or zero counts are found in the detector. The unbinned likelihood 
can be expressed as
\begin{equation}\label{likelihood2}
\log{\mathcal{L}}\propto -R_{T}+\sum_{k}^{N}{\log{R(t_{k},\omega_{k})}}
\end{equation}
where the sum runs over each of the $N$ detected counts and $R(t_{k},\omega_{k})$ is 
the event rate at the time $t_{k}$, energy $\omega_{K}$ of the $k$-event. $R_{T}$ is 
the expected number of counts over all exposure time and energy 
\begin{equation}
R(t,\omega)=B+S(t,\omega)
\end{equation}
where $B$ is the background rate of the detector and $S(t,\omega)$ is the expected rate from axions 
($\mathcal{N}_{\gamma}$), which depends on the axion properties $g_{a\gamma}$ and $m_{a}$ (see equation \ref{ngamma}).

The final analysis yields an upper limit on
$g_{ae}g_{a\gamma}\lesssim 8.1\times 10^{-23}\rm{GeV^{-1}}$
$(95\%~\rm{CL})$ for axion masses $m_{a}\leq 10\,\rm{meV}$ (see figure \ref{fig:gaegag}). Both
likelihood methods give compatible results on both, the best fit
value and the upper limit. In the latter case the deviation
of the limit remains within $\sim 0.3\%$ over the considered axion
mass range.

Parallel to the methods described above, a two-free parameter 
likelihood  was applied for axion masses within the reach of CAST. In this case, 
not only is the axion-electron coupling taken into account but also the Primakoff 
contribution to the axion production in the Sun in the frame of non-hadronic 
models. This approach computes a three-dimensional probability function at a given axion 
mass that correlates both $g_{a\gamma}$ and $g_{ae}$ couplings. 
The result yields a limit on the $g_{a\gamma}$--$g_{ae}$ 
parameter space (see figure \ref{fig:gaegag2}).

\subsection{Systematic uncertainties}
We studied the influence of systematic uncertainties on the best fit
value of $g_{ae}^{2}g_{a\gamma}^{2}$ and on the upper limit.
Statistically significant variations of the background on long and
short time scales are not apparent for the X-ray telescope
data~\cite{Kuster:2007ue}. Even if the background level were
time-dependent, it would not play a significant role, since the
X-ray telescope is measuring both, potential signal and background,
simultaneously. Since we observed no significant difference between
the background level and the spatial distribution during tracking
and non-tracking times, we used the same signal-spot area during
non-tracking periods to define the background. Alternatively, we
used tracking data and different regions on the CCD outside the
signal spot area to estimate the systematic uncertainties due to the
choice of background definition. The overall systematic uncertainty
is dominated by both the background definition and the pointing
accuracy, that affects the effective area of the telescope, the
location of the signal spot and its size, respectively. Other effects
such as uncertainties in the detector calibration, magnet parameters
and the likelihood method used are negligible in comparison with the
systematic induced by the choice of background definition. For the
best fit value of $g_{ae}^{2}g_{a\gamma}^{2}$ we find that in the
axion mass range for which CAST remains coherent we obtain
\begin{equation}
g_{ae}^{2}g_{a\gamma}^{2}\vert_{\rm{bestfit}}=\left(-1.136\pm_{2.46}^{3.09}{\rm stat.}\pm_{2.24}^{2.20}{\rm syst.}\right)\times 10^{-45}\,\rm{GeV^{-2}}\,.
\end{equation}

Alternatively, the background was also determined by extrapolating
the background measured during tracking periods in the part of the
CCD not containing the Sun spot. Different background selections led
to different upper limits on $g_{ae}^{2}g_{a\gamma}^{2}$, all of
them within the statistical uncertainty. 

\section{Conclusions}
Axions with tree-level coupling to electrons provide a different
physics case and phenomenology than hadronic models. From the
theoretical point of view, non-hadronic models are appealing because
they arise in grand unified theories (GUTs), well-motivated
completions of the standard model at high energies. From the
phenomenological side, it is worth noting that naturally the
coupling to electrons leads to larger axion fluxes from stars than
the coupling to photons. The sensitivity of CAST to non-hadronic
axions allows us to set a bound on the product of both coupling
constants $g_{ae}g_{a\gamma}\lesssim 8.1\times
10^{-23}~\rm{GeV^{-1}}$ for $m_{a}\leq 10\,\rm{meV}$.

For hadronic axions, Primakoff emission is the dominant production
process in stars and helioscope limits depend only on the
axion-photon interaction strength $g_{a\gamma}$. For low-mass axions,
the CAST limit on $g_{a\gamma}$ is competitive with, and even
somewhat superior to, the energy-loss limits from globular cluster
stars. For non-hadronic models, the stellar fluxes are dominated by
the BCA processes that are based on the axion-electron coupling
$g_{ae}$. In this case CAST constrains the product
$g_{ae}g_{a\gamma}$ in a significant way. However, the stellar
energy-loss limits on $g_{ae}$ are so restrictive here that CAST is
not yet competitive.

The claim of an anomalous energy loss in white dwarfs could be
an indication for the existence of axions with a coupling to
electrons in the $g_{ae}\sim 10^{-13}$ range, close to the
energy-loss limits of red giants in globular clusters. If we are to
assume this hint, the flux of solar axions is fixed by this
parameter and a next generation axion helioscope such as IAXO could
be able to detect it. We believe that the test of the white-dwarf
claim and surveying the DFSZ parameter space for the first time in a
laboratory experiment is a compelling motivation for IAXO and
strengthens its physics case.

In summary, there is a strong motivation to improve the helioscope
sensitivity beyond CAST down to $g_{a\gamma}\sim
10^{-12}~\rm{GeV^{-1}}$ and $g_{ae}\sim 10^{-13}$
\cite{Irastorza:2011gs}. This region includes, on the high-mass end,
a large set of favored QCD axion models, potentially supersedes
limits from SN~1987A and red giants in non-hadronic models, and
starts probing the parameters suggested by white-dwarf cooling.

\section*{Acknowledgments}

We thank CERN for hosting the experiment. This article has been authored by Lawrence 
Livermore National Security, LLC under Contract No. DE-AC52-07NA27344 with the U.S. Department 
of Energy. Accordingly, the United States Government retains and the publisher, by accepting 
the article for publication, acknowledges that the United States Government retains a 
non-exclusive, paid-up, irrevocable, world-wide license to publish or reproduce the published 
form of this article or allow others to do so, for United States Government purposes.
We acknowledge support from NSERC (Canada), MSES (Croatia) under Grant No.\ 098-0982887-2872, 
CEA (France), BMBF (Germany) under Grant Nos.\ 05 CC2EEA/9 and 05
CC1RD1/0, DFG (Germany) under Grant Nos.\ HO 1400/7-1 and EXC-153,
VIDMAN (Germany), GSRT (Greece), RFFR (Russia), the 
Spanish Ministry of Economy and Competitiveness (MINECO)
under Grant Nos.\ FPA2007-62833
and FPA2008-03456, NSF (USA) under Award No.\ 0239812, NASA under
Grant No.\ NAG5-10842, and the European Union under Grant No.\
PITN-GA-2011-289442 (ITN ``Invisibles''). J.~Redondo acknowledges support by the Alexander von 
Humboldt Foundation.

\end{document}